\documentclass[twocolumn,superscriptaddress,showpacs,preprintnumbers,prb,floatfix]{revtex4}
\usepackage{graphicx}
\usepackage{amsmath}
\usepackage{times}
\usepackage{epsfig}
\usepackage{multirow}
\usepackage{color}
\usepackage{xfrac}
\usepackage{cleveref}
\usepackage{gensymb}
\usepackage{doi}
\usepackage[english]{babel}
\begin{document}

\def\salto{\vskip 1cm} \def\lag{\langle} \def\rag{\rangle}

\newcommand{\redit}[1]{\textcolor{red}{#1}}
\newcommand{\blueit}[1]{\textcolor{blue}{#1}}
\newcommand{\magit}[1]{\textcolor{magenta}{#1}}

\newcommand{\SUNCAT} {SUNCAT Center for Interface Science and Catalysis, 
SLAC National Accelerator Laboratory and Department of Chemical Engineering
Stanford University,
443 via Ortega, Stanford, CA 94305, USA}

\title{Surface Energetics of Alkaline-Earth Metal Oxides:\\ 
Trends in Stability and Adsorption of Small Molecules.}

\author{Michal Bajdich}          \affiliation {\SUNCAT}
\author{Jens K. N\o rskov}       \affiliation {\SUNCAT}
\author{Aleksandra Vojvodic}     \affiliation {\SUNCAT}

\begin{abstract}
We present a systematic theoretical investigation of the surface properties, stability and reactivity, of rock-salt type alkaline-earth metal oxides including MgO, CaO, SrO, and BaO. The accuracy of commonly used exchange-correlation density functionals (LDA, PBE, RPBE, PBEsol, BEEF-vdW and hybrid HSE) and random-phase approximation (RPA) is evaluated and compared to existing experimental values.
Calculated surface energies of the four most stable surface facets under vacuum conditions: the (100) surface, the metal and oxygen terminated octopolar (111), and the (110) surfaces exhibit a monotonic increase in stability from MgO to BaO.
On the MgO(100) surface, adsorption of CO, NO, CH$_4$ is characterized by physisorption while H$_2$O chemisorbs, which is in agreement with experimental findings. 
We further use the on-top metal adsorption of CO and NO molecules to map out the surface energetics of each alkaline-earth metal oxide surface. 
The considered functionals all qualitatively predict similar adsorption energy trends. 
The ordering between the adsorption energies on different surface facets can be attributed to differences in the local geometrical surface structure and the electronic structure of the metal constituent of the alkaline-earth metal oxide. 
The striking observation that CO adsorption strength is weaker than NO adsorption on the (100) terraces as the period of the alkaline-earth metal in the oxide increases, is analyzed in detail in terms of charge redistribution within the $\sigma$ and $\pi$ channels of adsorbates.
Finally, we also present oxygen adsorption and oxygen vacancy formation energies in these oxide systems. 
\end{abstract}

\date{\today}

\maketitle
\section{Introduction}
Alkaline-earth metal oxides (AEMOs) of rock-salt structure
are simple ionic solids as opposed to more complicated transition metal oxides 
and their well-ordered surfaces are of great interest to theoretical and experimental surface science.~\cite{nygren_comparing_1996,freund_oxide_2008,freund_ultrathin_2008,mcfarland_catalysis_2013,campbell_enthalpies_2013} 
Due to their high stability and irreducibility, application of AEMOs in catalysis is limited to high temperature processes such as oxidative coupling of methane,~\cite{lunsford_katalytische_1995} but their activity can be enhanced via metal-doping,~\cite{mcfarland_catalysis_2013} or they can be utilized as support materials for other catalysts.~\cite{frondelius_charging_2008}
Furthermore, this computational catalyst study is intended
to fully explore and accurately map the surface properties such as stability, vacancy formation energetics, and adsorption of small molecules and bechmark theory to available experimental data.~\cite{campbell_enthalpies_2013}

Previous calculations of AEMO surfaces addressed the electronic structure, surface stability, 
relaxation, and rumpling effects of non-polar (100), (110), and (211) facets.~\cite{goniakowski_electronic_1994-1,goniakowski_relaxation_1995,broqvist_surface_2004,skorodumova_structural_2005} 
It was found that for a given facet, the surface energy decreases along the series going from MgO to BaO---an observation 
attributed mostly to a loss of Madelung energy, which is largest for MgO.~\cite{pacchioni_measures_1993,goniakowski_electronic_1994-1,goniakowski_relaxation_1995,broqvist_surface_2004} 
In addition, more open surfaces with greater number of cleaved bonds were found to be significantly 
less stable than the (100) surface due to a direct loss of energy associated with cleaved bonds. 
The origin of the positive rumpling for MgO and increasing negative rumpling for CaO, SrO, and BaO was explained on the grounds of a well-known narrowing of the O($2p$) valence band.~\cite{broqvist_surface_2004,skorodumova_structural_2005}
A stability study of the MgO(111) surface by Ciston {\it et.al.}~\cite{ciston_water-driven_2009} revealed the presence of stable octopolar terminations when no water was present and stable hydroxylated surface terminations in the presence of water. 
The MgO(100) surface is without a doubt the most studied surface of the considered AEMO series. In particular, adsorption of small 
molecules: CO, NO, CH$_4$, and H$_2$O has received much attention by both experiments~\cite{campbell_enthalpies_2013, wichtendahl_thermodesorption_1999,dohnalek_physisorption_2001,jung_dynamics_1991,tait_n-alkanes_2005,ferry_water_1998} and theory.~\cite{pacchioni_molecular_1992,nygren_bonding_1994,nygren_comparing_1996,valero_good_2008}
Important to mention are also recent accurate coupled-cluster benchmark studies of adsorption of CO~\cite{boese_accurate_2013} and CH$_4$.~\cite{tosoni_accurate_2010} 
Previous theoretical studies of adsorption of water on the (100) surfaces of the AEMO series 
revealed the preference towards dissociation of the water molecules for CaO to BaO.~\cite{carrasco_dynamic_2008}
Perhaps the least studied surface properties are related to oxygen adsorption and oxygen vacancy formation. To our knowledge only a handful of 
studies exist for oxygen vacancy formation on the MgO(100) surface.~\cite{kappers_f+_1970,ferrari_electronic_1995,richter_concentration_2013}

In this paper, we present a systematic computational study of surface properties of four AEMOs: MgO, CaO, SrO, and BaO, 
by mapping their surface energies, CO, NO, and oxygen adsorption energies, and oxygen vacancy formation energies of 
the most stable surfaces of these oxides. While the surface energy is a determining factor for equilibrium morphology, CO and NO adsorption energies serve as probe molecules of surface reactivity.~\cite{lamberti_probing_2010}
Another important surface chemistry property of an oxide surface is its oxygen chemistry, {\it i.e.}, a measure of the interaction between the surface and atomic oxygen on one hand and oxygen vacancies on the other hand, since these play a key role as active centers in most oxidative and dehydrogenation reactions processes.~\cite{mcfarland_catalysis_2013}
From a practical point of view, it is highly desirable that a computational catalyst study achieves an accurate mapping of possible surface energetics. In order to benchmark our theoretical approaches, we evaluated the accuracy of four levels of density functional theory (DFT) theory: the local density approximation (LDA), the generalized-gradient approximation (GGA), van der Waals corrected GGA (GGA+vdW), screened hybrid-GGA as well as post-DFT random-phase approximation (RPA), and assessed our findings against existing experimental data.

The remainder of this paper is organized as follows. First we give a brief introduction of our computational methodology, followed by a presentation of the results for bulk properties and surface stabilities of the four considered AEMOs. Next, we discuss the performance of common DFT functionals and the RPA method for adsorption of CO, NO, CH$_4$, and H$_2$O molecules on the MgO(100) surface. Then a mapping of the CO and NO adsorption energies on the different surfaces of the AEMOs is discussed. 
Additionally, we also provide a detailed analysis of the bond character for representative systems, including a comparison between CO@MgO(100) and CO@BaO(100), and between NO@MgO(100) and NO@BaO(100). In the last part of the Results section, we discuss the oxygen chemistry of the investigated surfaces. Finally, we conclude with a summary of our most relevant findings and their implications.

\section{Methods} 
We employed the periodic plane-wave basis set implementation of DFT within the PWscf program of Quantum Espresso.~\cite{giannozzi_quantum_2009} 
We performed calculations with several DFT functionals frequently used in solid-state and surface science studies: LDA,~\cite{perdew_accurate_1992} GGA: PBE,~\cite{perdew_generalized_1996,perdew_generalized_1996-1} RPBE,~\cite{hammer_improved_1999} and PBEsol,~\cite{perdew_restoring_2008} GGA with van der Waals (vdW) density functional: BEEF-vdW,~\cite{wellendorff_density_2012} and screened GGA-hybrid: HSE.~\cite{heyd_hybrid_2003,heyd_erratum:_2006} 
The BEEF-vdW approach combines semilocal Bayesian error estimation functional with an additional vdW nonlocal correlation term.
The vdW term is computed at via DFT/vdW-WF2 method,~\cite{hult_density_1996,silvestrelli_van_2009,lee_higher-accuracy_2010} which been shown to lead to improved 
description of surface processes.~\cite{tonigold_adsorption_2010,silvestrelli_adsorption_2012,silvestrelli_inclusion_2013}  
For elements from H to Mg we used the PBE generated norm-conserving pseudopotentials of Trouiller-Martins~\cite{fuchs_ab_1999,troullier_efficient_1991} while for Ca, Sr, and Ba we used the pseudopotentials of Goedecker--Hartwigse--Hutter--Teter,~\cite{hartwigsen_relativistic_1998} which explicitly include semicore states. 
The use of norm-conserving pseudopotentials was dictated by the implemenation of GGA-hybrid part of PWscf. 
A kinetic-energy cutoff of 80~Ry and 4 times that value for the charge density cutoff was used in the calculations except for the RPA calculations as described below.

The random-phase approximation~\cite{bohm_collective_1951,pines_collective_1952,bohm_collective_1953} calculations where performed using the PAW method\cite{blochl_projector_1994} within the plane-wave VASP code.~\cite{kresse_ab_1993,kresse_efficiency_1996,kresse_ultrasoft_1999}  
The RPA energy at PBE optimized geometries were evaluated either i) in conventional RPA@PBE scheme, where the exchange and correlation are  
calculated non-selfconsitently from PBE orbitals as $E_{RPA@PBE}=E^{ex}_{@PBE}+E^{corr}_{@PBE}$, or ii) in hybrid RPA$_h$@PBE scheme,\cite{ren_beyond_2011} 
where the effect of self-consistent exchange energy is added as $E_{RPA_h@PBE}=E^{ex}_{@HF}+E^{corr}_{@PBE}$. 
The convergence of energy differences within few meVs was achieved with plane-wave cutoff of 350 eV for GW variants of VASP PAW potentials. 
Additional computational details are provided in connection to the rest of the results.

\section{Results}~\label{results}
\subsection{Bulk properties}
\begin{figure}[!ht]
\includegraphics[width=\columnwidth]{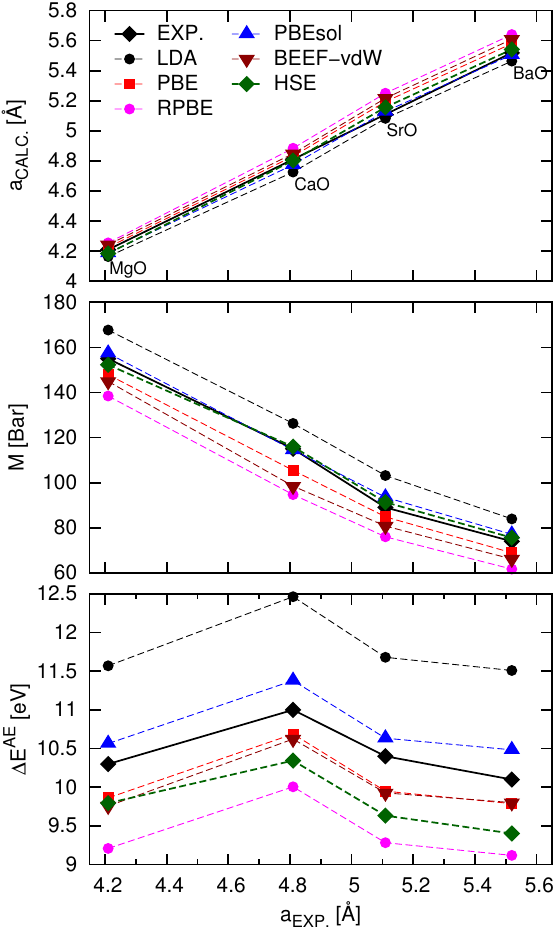}
\caption{Calculated lattice constants $a$, bulk moduli $M$ and atomization energies $\Delta E^{\rm AE}$ for six different DFT functionals together with experimental values adapted from Refs.~[\onlinecite{chang_pressure_1969,chang_elastic_1977,konigstein_ab_1998}]. Note, that in our simplified comparison to experimental data, we have neglected anharmonic ZPE corrections, which can be as large as $\Delta a= -0.02$~\AA\, for bulk MgO.~\cite{schimka_improved_2011} 
}\label{fig:bulk}
\end{figure}

The calculated bulk properties: lattice constants $a$, bulk moduli $M$, and atomization energies $\Delta E^{AE}$ of the AEMO series are presented in Fig.~\ref{fig:bulk}. 
For each of the observables, our calculated values follow the experimentally observed trend~\cite{chang_pressure_1969,chang_elastic_1977,konigstein_ab_1998}
and are also in agreement with previous theoretical findings.~\cite{broqvist_surface_2004,skorodumova_structural_2005}
We find an increase in the lattice constant and a decrease of the bulk modulus going from MgO to BaO within the AEMO series, which is
consistent with the increase of the ionic radii of the metal ions and shows a transition from ionic to more covalent bonding.~\cite{pacchioni_measures_1993} 
The atomization energy exhibits a maximum for CaO but the variations between the different AEMOs is fairly constant within a 1~eV range for a given DFT functional.

There are also characteristic trends in the above calculated quantities for a given DFT functional and between different functionals. 
The well-known overestimation of the stability within LDA is clearly visible for AEMOs. The lattice constants are too short and bulk moduli 
and atomization energies are too large. 
On the other hand, the RPBE predictions are on the opposite side of the spectrum, with too long lattice constants and underestimation of atomization energies and bulk moduli. 
All other functionals included in this study lie within the LDA--RPBE bounds. A comparison of our PBE, PBEsol, and HSE data with experimental data for bulk MgO show the following lattice constant dependence $a_{\rm PBE}>a_{\rm PBEsol}\gtrsim a_{\rm HSE}\gtrsim a_{\rm EXP}$ (and similarly for bulk moduli $M$)
while the atomization energy follows $\Delta E^{\rm AE}_{\rm PBE}\approx \Delta E^{\rm AE}_{\rm HSE}< \Delta E^{\rm AE}_{\rm EXP} < \Delta E^{\rm AE}_{\rm PBEsol}$. These findings are in qualitatively agreement with the ones of Shimka \textit{et al.}~\cite{schimka_improved_2011}
We find that the bulk properties obtained with the BEEF-vdW functional are not significantly improved over values obtained by the PBE functional.  
When it comes to the band gaps, the calculated values within the HSE functional (MgO: 6.7~eV, CaO: 5.3~eV; SrO~eV: 4.7~eV; BaO: 3.2~eV) are, as expected, the closest to experimental estimates (7.8~eV, 7.1~eV, 5.9~eV, and 4.3~eV from Ref.~[\onlinecite{rao_logarithmic_1979}]).
For all functionals, the calculated density of states (DOS) of the bulk AEMOs (not shown) exhibit a narrowing of the valence O($2p$) band and a shift of the metal M($p$) semicore states towards higher energies within the series from MgO to BaO in agreement with previous studies.~\cite{broqvist_surface_2004,skorodumova_structural_2005}

\subsection{Energetics of Stoichiometric Surfaces}\label{sec:surfe}
\begin{figure}
\includegraphics[width=\columnwidth]{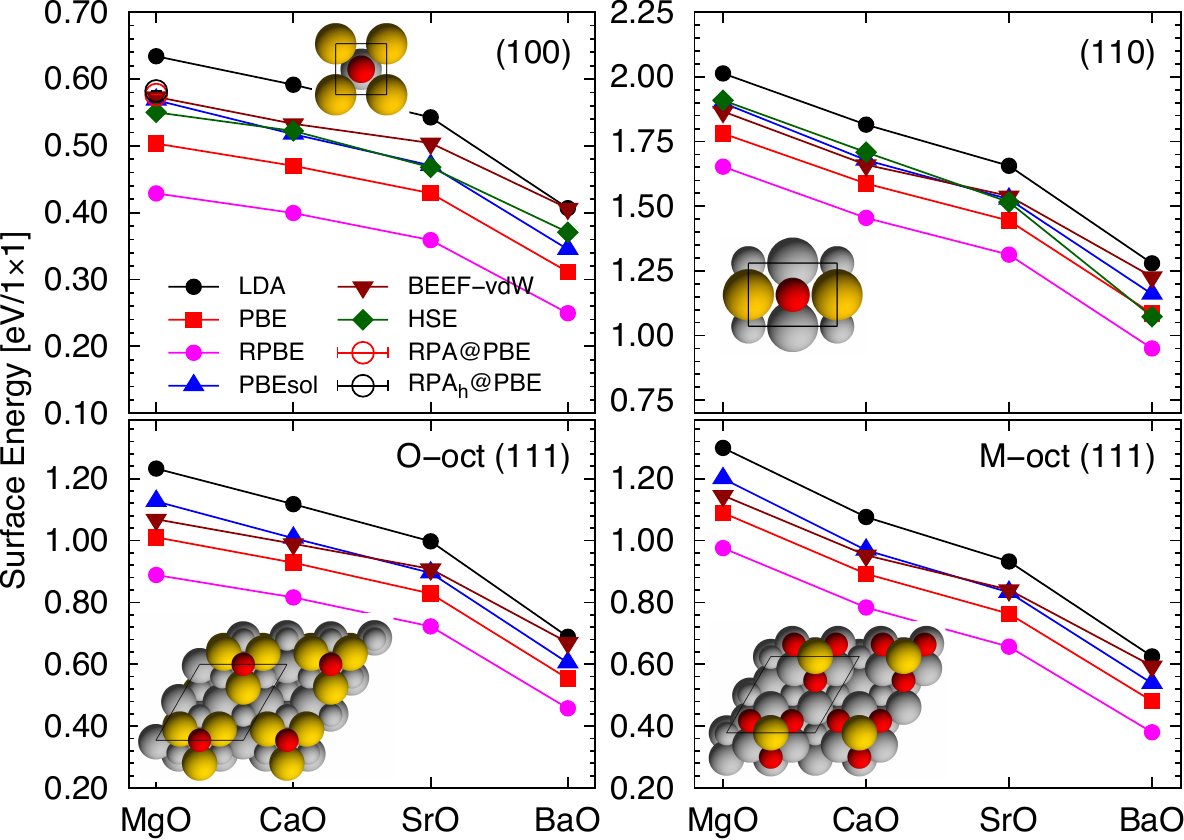}
\caption{Calculated surface energies of stoichiometric low-index terminated alkaline-earth metal-oxide surfaces in eV normalized to 
$1\!\times\!1$ surface area. The results for the (100), (110), and two octopolar (111) terminated (M-Oct, O-oct) surfaces obtained using 
six different DFT functionals are shown. 
The corresponding primitive cell of the surfaces used in the calculations are also shown as insets, where larger spheres represent the metal atoms. For clarity, only the top layer was rendered in color.}\label{fig:surface_energy}
\end{figure}

The most stable surface termination of rock-salt type compounds is the non-polar (100) surface, 
which has been argued to be due to its close packed bulk-like structure.~\cite{tasker_stability_1979} 
In terms of stability, the (110) surface followed by the (211) surface are other low-index stoichiometric non-polar surface terminations characterized by 
increasing number of uncoordinated bonds at the surface.~\cite{goniakowski_relaxation_1995} The polar (111) surface 
is known to reconstruct to form two different octopolar terminations~\cite{wolf_reconstruction_1992,wolf_structure_1995} (metal and oxygen terminated: M-oct and O-oct) with a $2\times2$ periodicity.  Additionally, in the presence of water, the (111) facet can be greatly stabilized by hydroxilation of the surface.~\cite{ciston_water-driven_2009} 

For this study, we have selected four low-index stoichiometric surface terminations under vacuum conditions namely the (100), (110) 
and the M-oct and O-oct (111) surfaces.
Their calculated surface energies normalized to $1\!\times\!1$ unit surface area are summarized in Fig.~\ref{fig:surface_energy}. 
The surface energy is obtained via the linearized method of Fiorentini and Methfessel.~\cite{fiorentini_extracting_1996} Slabs from 2 to 8 metal-oxide layers have been studied with a surface energy convergence within 0.01~eV, which was achieved for 4-6 layers. 
The periodic symmetric surface slabs were separated by 16~\AA\, of vacuum, cleaved at bulk geometries and fully relaxed within fixed unit cells bellow maximum threshold force of 0.05~eV/{\AA}.
To be noted is that the HSE investigation of the (111) surfaces was omitted due to prohibitively large computational cost within plane-wave PWscf for the larger slabs. 

For all surfaces, we find that the calculated surface energy decreases monotonically along the series from MgO to BaO, as 
has been attributed previously to a loss of Madelung energy, which is largest for MgO.~\cite{pacchioni_measures_1993,goniakowski_electronic_1994-1,goniakowski_relaxation_1995,broqvist_surface_2004} 
Furthermore, the energy differences between the facets also decrease along the series. 
This observation has strong implications for the surface morphology of AEMOs, with BaO having much larger probability 
of having different facets exposed than MgO, as first pointed out by Broqvist \textit{et al}.~\cite{broqvist_surface_2004}
We also find, that the surface energies of the two octopolar (111) terminations fall in-between the (100) and (110) surfaces. 
For MgO, the O-oct (111) surface has a lower surface energy relative to the M-Oct (111) surface, as observed by Ciston.~{\it et. al}.\cite{ciston_water-driven_2009}
Interestingly, the stability between the O-oct and M-Oct (111) surfaces is reversed for the rest of the AEMO series (see lower panels of Fig.~\ref{fig:surface_energy}). 
For a given metal-oxide, the ordering of stabilities between the individual surfaces approximately corresponds to coordination loss 
per surface metal site, which is 1/6 for (100), 1.5/6 on average for (111) M-Oct (same for  O-site in O-oct termination) and 2/6 for (110).

\begin{table}
\caption{ MgO(100) surface energies $\sigma$ in eV per $1\!\times\!1$ area compared to experiment.
\label{tab:table1}}
\begin{tabular}{l c c c c c c c c c}
\hline
\hline 
Method & EXP.~\cite{jura_experimental_1952} & LDA & PBE & RPBE & PBEsol   \\
\hline
$\sigma$[eV/$1\!\times\!1$] & 0.575 & 0.634 &  0.503 & 0.429 & 0.568 \\
&  & BEEF-vdW & HSE & RPA@PBE & RPA$_h$@PBE \\
$\sigma$[eV/$1\!\times\!1$] &  & 0.573 & 0.550 & 0.577 & 0.582 \\
\hline  
\hline
\end{tabular}
\end{table}
Qualitatively, each of the investigated DFT functionals predicts similar trends for surface stability of the AEMO series. 
We find that the surface energies lie within the bounds of LDA (too high) and the RPBE (too low), as we found for bulk properties.
For MgO(100), the calculated surface energies are shown in Table~\ref{tab:table1}, 
with PBEsol, BEEF-vdW, and RPA methods having the best agreement with the experimental finding.
As experimental reference, we have chosen the value of 0.575~eV,~\cite{jura_experimental_1952} which is based on thermodynamic 
measurements of polycrystaline samples as opposed to experiments based on cleavage energy of single crystals,\cite{westwood_cleavage_1963} which are 
likely to overestimate this quantity.\cite{ciston_why_2010} The good performace of PBEsol functional is expected, given its design to match the jelium surface energies.~\cite{armiento_functional_2005,csonka_assessing_2009}  Contrary to bulk properties, the improved performance of the BEEF-vdW functional over the PBE functional for surface energies is somewhat surprising and suggests the importance of long-range interactions for surface properties.

\subsection{Adsorption on the MgO(100) Surface}
\begin{figure}
\includegraphics[width=\columnwidth]{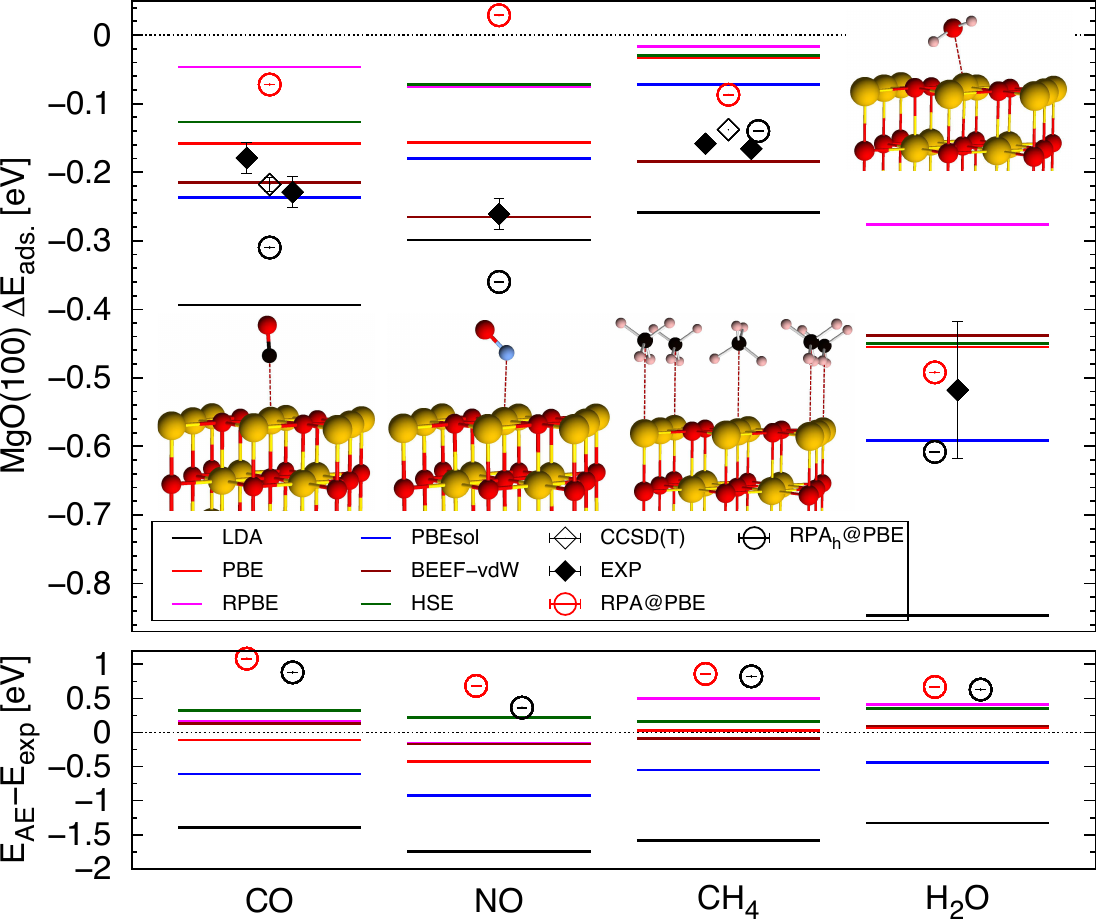}
\caption{Upper panel: Adsorption energies of CO, NO, CH$_4$, and H$_2$O molecules on the MgO(100) surface.
The $2\!\times\!2$ unit cells are shown as insets with geometries discussed in the text.
Except for the case of CH$_4$ with one monolayer coverage, 
all other adsorbates are in the limit of low coverage (up to coverage of $\Theta=1/8$) as discussed in text. The experimental and calculated values are 
summarized in Table~\ref{tab:table2}. 
Lower panel: Associated atomization errors of each molecular species for the different methods
relative to the ZPE corrected experimental atomization energies as adapted from NIST CCCBD.~\cite{edited_by_russell_d._johnson_iii_nist_2013}}\label{fig:mgo100}
\end{figure}

\begin{table}
\renewcommand{\thefootnote}{\alph{footnote}}
\renewcommand{\thempfootnote}{\alph{mpfootnote}}
\caption{Adsorption energies on the MgO(100) surface in eV.
\label{tab:table2}}
\begin{tabular}{l c c c c}
\hline
\hline 
Method & CO & NO & CH$_4$ & H$_2$O\\
\hline
EXP    & -0.18(2)~\cite{wichtendahl_thermodesorption_1999}   &  -0.26(2)~\cite{{wichtendahl_thermodesorption_1999}}  &  -0.16(1)~\cite{jung_dynamics_1991}   &   -0.52(10)\cite{ferry_water_1998} \\
       & -0.23(2)~\cite{dohnalek_physisorption_2001}   &   -                        &  -0.17(1)\cite{tait_n-alkanes_2005}   &    -     \\
CCSD(T)  & -0.218~\cite{boese_accurate_2013}                 &   -    &  -0.138~\cite{tosoni_accurate_2010}    &    -       \\ 
\hline
LDA    & -0.393  & -0.299  &  -0.259  &  -0.847 \\
PBE    & -0.158  & -0.157   &  -0.033  &  -0.455 \\
RPBE   & -0.046  & -0.076 &  -0.016 &  -0.276 \\
PBEsol & -0.237  & -0.180  &  -0.072 &  -0.592 \\
BEEF-vdW   & -0.215  & -0.266  &  -0.184  &  -0.438 \\
HSE    & -0.127    & -0.072   &  -0.030  &  -0.450 \\
RPA@PBE    & -0.072     &  0.029     &  -0.087     &  -0.492    \\
RPA$_h$@PBE   & -0.310      & -0.360      &  -0.140      &  -0.608    \\
\hline
\hline
\end{tabular}
\end{table}

The MgO(100) surface is the most studied surface of the AEMO series and its interaction with different adsorbates has been experimentally well characterized (see \textit{e.g.} review of Campbel~\textit{et al.}~\cite{campbell_enthalpies_2013}).
For direct comparison with experiments~\cite{wichtendahl_thermodesorption_1999,dohnalek_physisorption_2001,jung_dynamics_1991,tait_n-alkanes_2005} and known theoretical benchmarks,~\cite{pacchioni_molecular_1992,nygren_comparing_1996,carrasco_dynamic_2008,valero_good_2008,boese_accurate_2013,tosoni_accurate_2010}  we have chosen to study adsorption of CO, NO, and H$_2$O at low adsorbate coverage as well as CH$_4$ at what is often referred to as one monolayer coverage (50\% occupancy of metal sites).

Figure~\ref{fig:mgo100} summarizes the calculated adsorption energies, which have been obtained 
from slab models consisting of four MgO layers, 
where the bottom 2 layers are fixed to bulk positions and the remaining top 2 layers are fully relaxed.  
The slabs were separated with a 16~{\AA} of vacuum. 
The adsorption energies have been extrapolated from $\sqrt{2}\!\times\!\sqrt{2}$ to $\sqrt{8}\!\times\!\sqrt{8}$ cells, 
with values converged within 0.01~eV at intermediate $2\!\times\!2$ cells. Due to higher computational cost, the reported values for HSE and RPA calculations are only for intermediate cell sizes. 

The geometries of physisorbed CO and NO molecules on the MgO(100) surface at low coverage are relatively well known from previous theoretical studies.~\cite{pacchioni_molecular_1992,nygren_comparing_1996,valero_good_2008,boese_accurate_2013} 
At this coverage, the CO molecule adsorbs on-top of the Mg metal site with a C-Mg bond length of approximately $2.4$~{\AA}
and a small tilting angle of $5\degree$. In contrast to CO, the extra unpaired electron of the NO leads to a strongly tilted adsorption geometry of the NO molecule ($50\pm10\degree$) with approximately the same N-Mg bond length of $2.4$~{\AA} as for the C-Mg bond of an adsorbed CO molecule.
Experimentally measured adsorption energies of both CO and NO find a weak binding to the surface (see Fig.~\ref{fig:mgo100}), likely with a long range dispersive character of the bond mediated by dipole interactions. Importantly, NO adsorbs marginally stronger than CO by approximately 1$\pm$ 0.5 kcalmol$^{-1}$. We also note, that both experimental estimates have been extrapolated to zero coverage and include zero point vibrational energy of -4 kJmol$^{-1}$ as adapted from Ref.~[\onlinecite{boese_accurate_2013}].

In Fig.~\ref{fig:mgo100} we compare the adsorption energies predicted from six DFT functionals to experimental results for CO,~\cite{wichtendahl_thermodesorption_1999,dohnalek_physisorption_2001} and NO,~\cite{wichtendahl_thermodesorption_1999}
as well as to available accurate quantum chemical CCSD(T) adsorption benchmarks.~\cite{boese_accurate_2013}
We find that all calculated and experimental adsorption energies lie within the bounds of LDA (too strong adsorbate-surface interaction) and RPBE (too weak adsorbate-surface interaction), as was observed for the bulk and surface properties.
However, LDA and PBEsol and to lesser degree also the PBE functional predict that CO adsorbs stronger than NO (see Fig.~\ref{fig:mgo100}). 
One possible origin of this discrepancy could lie in the larger overbinding error found for the gas-phase NO as compared to CO molecule that have been found for these functionals (see atomization errors of lower panel of Fig.~\ref{fig:mgo100}). 
The experimentally observed adsorption trend is accurately captured within RPBE, HSE, and BEEF-vdW. 
We find that the BEEF-vdW extrapolated adsorption energies for CO and NO are in best agreement with experimental values, 
while HSE predicts slightly too weak adsorption suggesting the importance of the long-range interactions for these systems.
Despite that RPBE predicts the correct order between CO and NO adsorption energies, 
the adsorbate-surface bond strength is too weak---which can not be improved by inclusion of vdW interactions (not shown). 

When it comes to adsorption of CH$_4$, the most favorable coverage is one monolayer, {\it i.e.}, one methane molecule on top of every other metal atom in what is known as the ``dipod'' configuration (two hydrogen bonds pointing upwards and two hydrogen bonds pointing downwards towards the nearest surface oxygen atoms).~\cite{jung_dynamics_1991,tait_n-alkanes_2005}  
Two dipods rotated $90\degree$ relative to each other form the most stable $2\!\times\!2$ periodic superstructure,~\cite{tosoni_accurate_2010} which was also employed in this study. 
The interaction between the CH$_4$ molecule and the MgO(100) surface has a predominantly vdW nature with small or no contribution from dipole interactions,~\cite{tosoni_accurate_2010} and this is the least bound adsorbate of all studied systems.
We find that from all DFT functionals, only the BEEF-vdW predicted adsorption energy lies in the vicinity 
of the experiment and CCSD(T) benchmark. The obtained BEEF-vdW average equilibrium bond distance between the CH$_4$ molecules and the surface is $3.32$~{\AA}, which agrees with the experimental value of $3.30$~{\AA}.~\cite{robert_m._hazen_effects_1976}   

Adsorption of H$_2$O on MgO(100) (Fig.~\ref{fig:mgo100}) is an example of a weakly chemisorbed system with significant hydrogen to surface bonding.~\cite{ferry_water_1998} The lowest energy structure is best described as water lying flat on the surface with one hydrogen pointing towards the surface oxygen.~\cite{carrasco_dynamic_2008}
The PBE, PBEsol, BEEF-vdW, and HSE calculated adsorption energies of water on the MgO(100) surface all lie within the
relatively broad experimental estimate. 

Finally, we would like to comment on the performace of the RPA method.
In Fig.~\ref{fig:mgo100} we compare the performance of the conventional RPA scheme as well as the hybrid RPA scheme.
Clearly, the hybrid RPA outperforms the conventional RPA scheme for CO, NO, and CH$_4$ adsorption and predicts 
experimental ordering of energies, albeit at slightly overestimated values for CO and NO when compared to experiments. 
On the other hand, the RPA@PBE scheme predicts too weak adsorption bonds and has difficulty with describing the NO adsorption, which is the only spin-polarized system of the ones investigated here.
For H$_2$O adsorption, both methods deliver comparable results. Lastly, RPA$_h$@PBE also reduces the well-known atomization errors of conventional RPA (lower panel of Fig.~\ref{fig:mgo100}). 

\subsection{CO and NO Adsorption in the AEMO Series}
\begin{figure}
\includegraphics[width=\columnwidth]{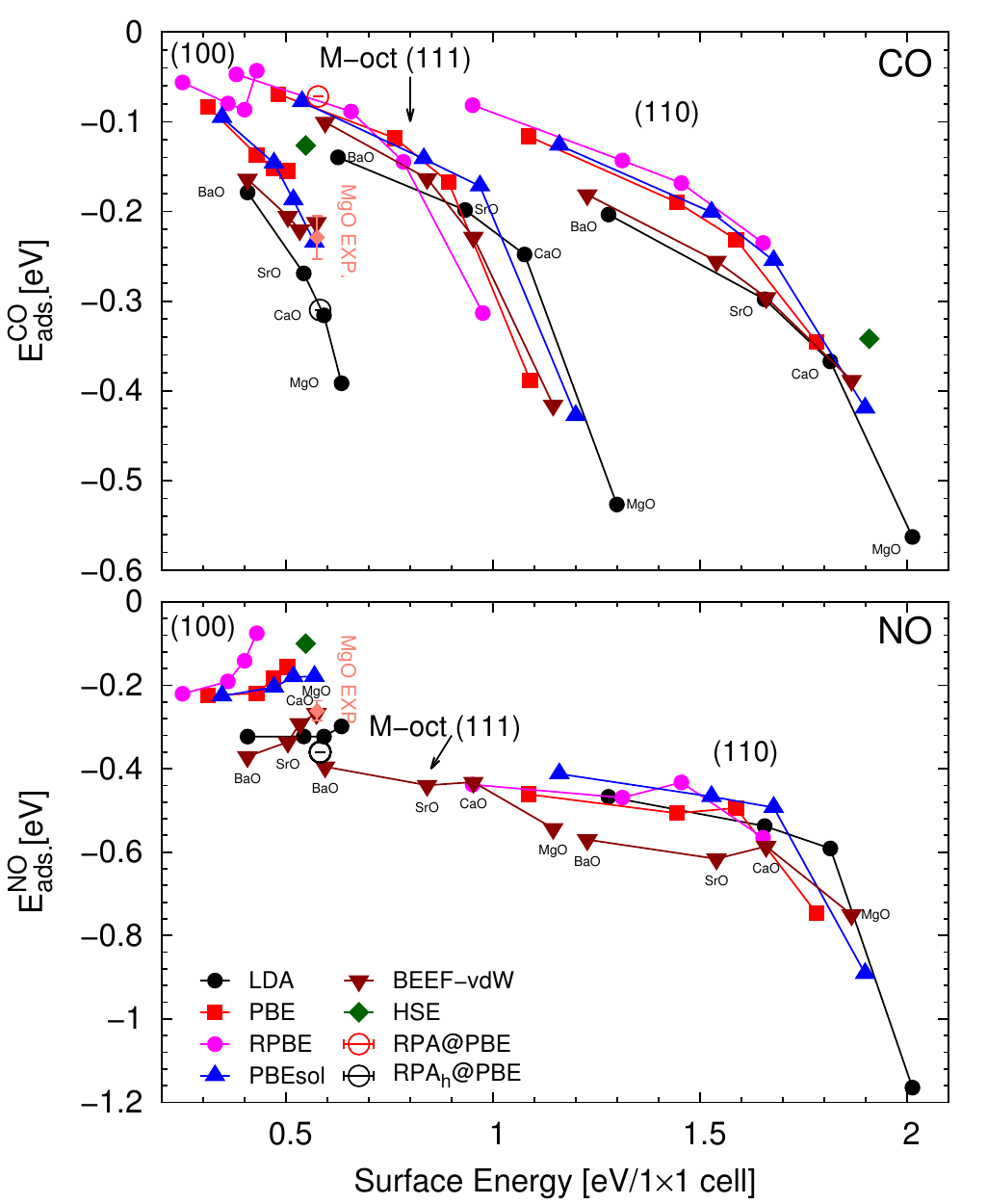}\\
\caption{Adsorption energies of CO (top panel) and NO (bottom panel) in the on-top surface metal site on the (100), (111)-M-oct, and (110) surfaces
of the AEMO series plotted as function of surface energy. For clarity, only the BEEF-vdW results are shown for NO on (111)-M-oct.}\label{fig:COads}
\end{figure}

In this section, we focus on the adsorption energy trends of CO and NO on the most stable surface terminations of the different AEMOs.
The adsorption energies of CO and NO molecules in the on-top surface metal site as a function of the surface energy, which was shown to be a monotonic function of the AEMO series (see Fig.~\ref{fig:surface_energy}), are presented in Fig.~\ref{fig:COads}. 
As for the MgO(100) surface, we employed 4 layer slab models with the bottom 2 atomic layers fixed in bulk positions and remaining top 2 layers fully relaxed for all surfaces. 
All the energetics is calculated using $2\!\times\!2$ simulation cells, corresponding to a coverage of $\Theta=0.25$ for the (100), (110), and with only single on-top site for (111)-M-oct surface. The on-top metal site of the (111)-O-oct surface is occupied by oxygen and does not allow for additional adsorption, therefore it is omitted in the rest of our analysis. 

The most general trend (see Fig.~\ref{fig:COads}) is that for a given surface termination, 
both the $E_{\rm ads.}^{\rm CO}$ and the $E_{\rm ads.}^{\rm NO}$ get progressively weaker as we move from MgO to BaO within the AEMO series. 
The only exception is the case of NO adsorbed on the (100) terraces, where we observe a
strengthening of the NO to surface bond. 
In addition, we find that for both CO and NO on a given metal oxide, a less stable surface on average adsorbs a molecule stronger than a more stable surface. This behavior is expected, as less stable surfaces are also more open, {\it i.e.}, are characterized by a larger number of uncoordinated bonds,~\cite{tasker_stability_1979,goniakowski_relaxation_1995} which in turn can interact with the adsorbate. When it comes to the surfaces studied here, we observe that the (100) surface has 5-fold coordinated metal sites, the (110) surface has 4-fold  
coordinated metal sites, and the (111)-M-Oct surface has single 3-fold and triple 5-fold coordinated metal sites.
As we have discussed in Section~\ref{sec:surfe}, the above average number of uncoordinated bonds is a relatively good predictor of the surface energy. However, for the case of CO and NO adsorption (see Fig.~\ref{fig:COads}), we find that there is no significant difference between 
the adsorption energies on the (110) and (111)-M-Oct surface, while there is a clear difference between the (100) surface and the other two surface terminations. 
Hence, we conclude that bond counting or the use of surface energy to predict the adsorption energetics is limited and provides only a rough estimate 
of the reactivity of different surfaces. In other words, they can not be used as descriptors since they are unable to give any quantitative comparison 
nor provide an ordering between the different oxide surfaces.

We also find that NO always binds stronger to the surface than CO. 
This is the same dependence as for the experimental results on the MgO(100) surface as we discussed in the previous section. 
We also note that the same ordering of CO vs NO adsorption energies is found on the NiO(100) surface.~\cite{valero_density_2010}    
In the next section, we analyze in greater detail the origin of this effect. 

\subsection{Electronic Structure Analysis of CO and NO Bond Formation}
\begin{figure}
\begin{tabular}{c c}
\hspace{-5pt} \includegraphics[height=0.5\columnwidth]{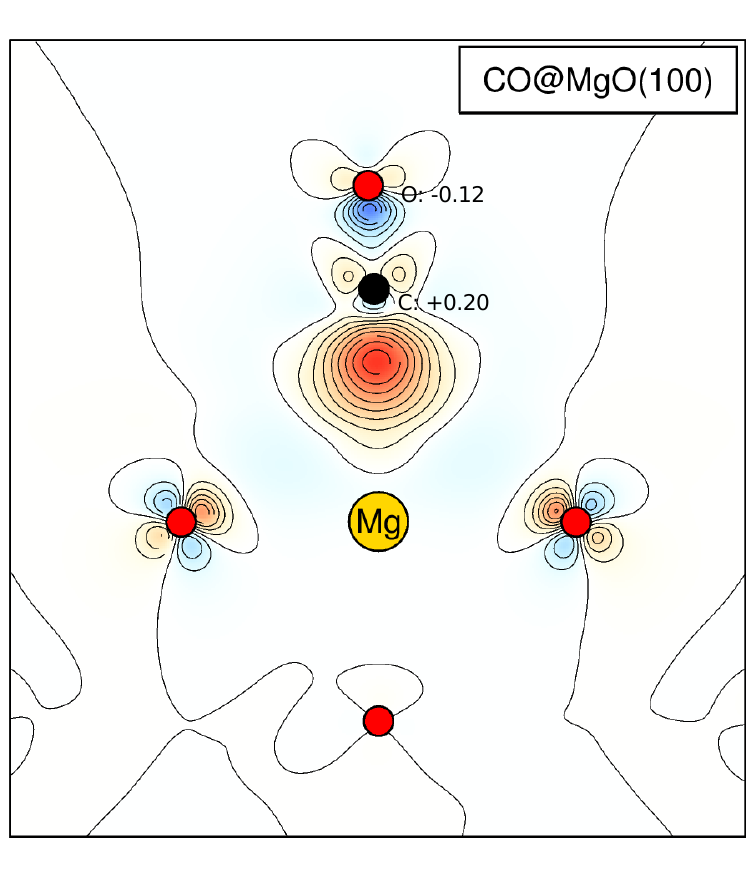} & 
\hspace{-5pt} \includegraphics[height=0.5\columnwidth]{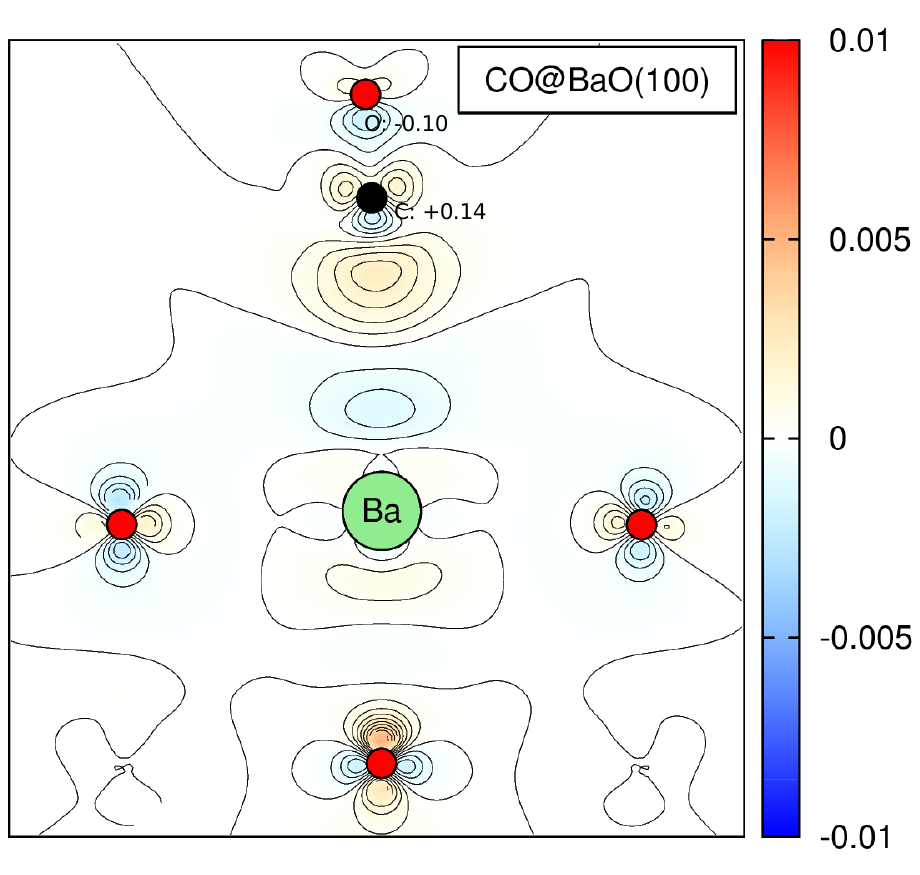} \hspace{-5pt} \\
\end{tabular}
\caption{Charge density difference plots for CO@MgO(100) (left) and CO@BaO(100) (right) projected along the $\langle110\rangle$ direction 
obtained with the BEEF-vdW functional. The positive (negative) values of this quantity indicate regions with gain (loss) of electronic charge.
The numeric labels indicate the change in the Bader charge upon adsorption relative to the free gas-phase molecule.  
Small red and black spheres indicate positions of oxygen and carbon atoms, respectively.}\label{fig:COddplot}
\end{figure}
\begin{figure}
\begin{tabular}{c c}
\hspace{-5pt} \includegraphics[height=0.55\columnwidth]{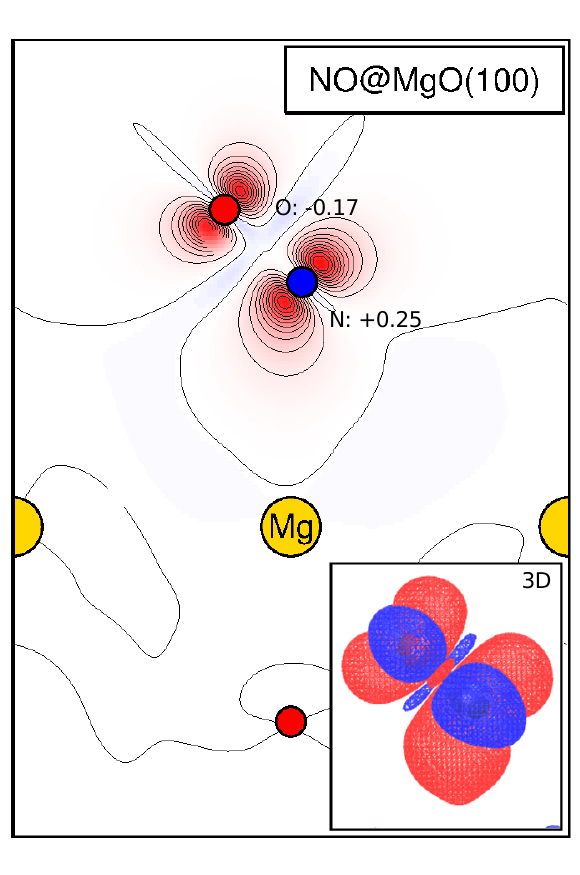} & 
\hspace{-5pt} \includegraphics[height=0.55\columnwidth]{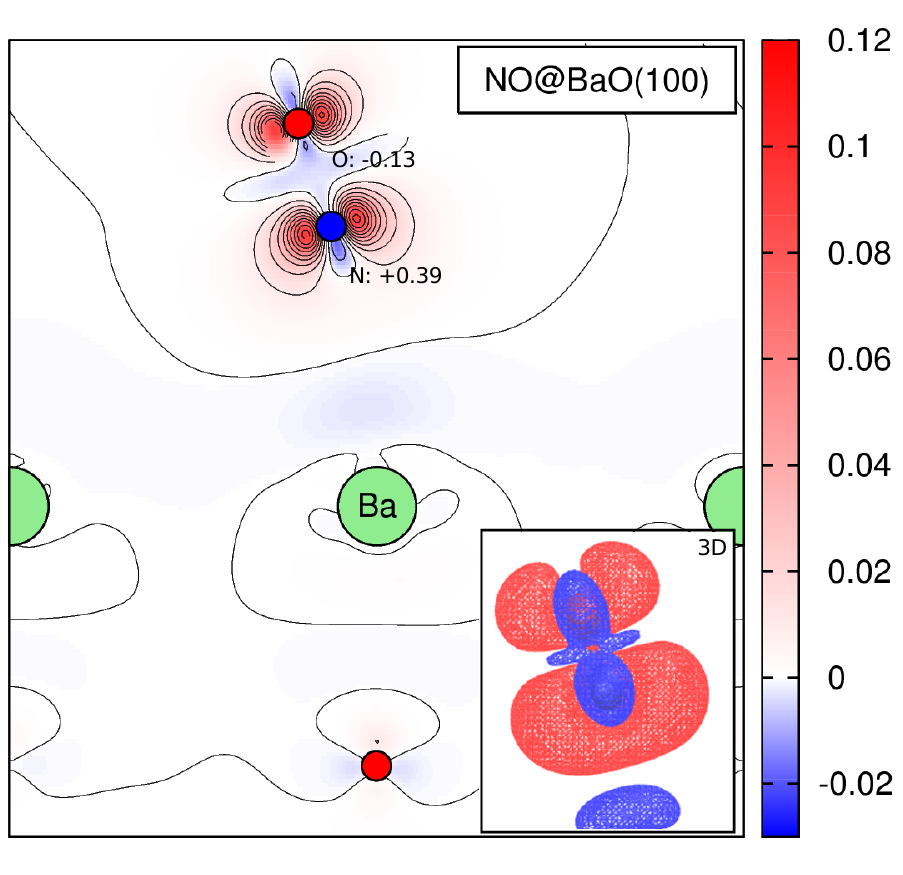} \hspace{-5pt} \\
\end{tabular}
\caption{Charge density difference contour plots for NO@MgO(100) (left) and NO@BaO(100) (right) projected along the $\langle100\rangle$ direction 
calculated at the BEEF-vdW level. Small blue sphere indicate positions of the nitrogen atom while 
the rest of the labels are identical to Fig.~\ref{fig:COddplot}.
Insets: Full three-dimensional shapes of charge density difference contours at values of $\pm$0.001.}\label{fig:NOddplot}
\end{figure}

Here we discuss the above mentioned observation that CO adsorbs marginally ($\sim$0.05 eV) weaker on BaO(100) 
than on MgO(100), while NO adsorbs substantially stronger ($\sim$0.1 eV) on BaO(100) than on MgO(100). 
The analysis is based on representative results obtained with the BEEF-vdW functional. We have chosen the BEEF-vdW functional 
due to its predictability of the CO and NO adsorption on the MgO(100) surface.
In order to understand the reasons behind differences in chemical bonding, it is instructive to plot the charge density difference (CDD) upon adsorption.
This quantity is obtained as $\rho_{\rm CDD}=\rho_{\{\rm Ads.+surf.\}}-\rho_{\rm Ads.}-\rho_{\rm surf.}$, where $\rho_{\rm Ads.}$, $\rho_{\rm surf.}$ and  $\rho_{\{\rm Ads.+surf.\}}$ are self-consistent densities of the adsorbate, surface and of the combined system, respectively. We note that $\rho_{\rm Ads.}$ and  $\rho_{\rm surf.}$ have been calculated based on the fixed geometry of the combined system to allow for identification of true electronic effects and avoid geometric effects. 
 
Figures~\ref{fig:COddplot} and~\ref{fig:NOddplot} show the CDDs for CO and NO on the (100) surface of MgO and BaO.
For the adsorbed CO molecule, $\rho_{\rm CDD}$ depicts polarization in the $\sigma$ channel, which is about twice as large for MgO than for BaO. This is also quantitatively supported from the calculated Bader charges as indicated in Figs~\ref{fig:COddplot} and~\ref{fig:NOddplot}. 
For the adsorbed NO molecule, on the other hand, the overall change in the density due to adsorption is much larger than the one for CO. 
Both CDD plots for adsorbed NO are dominated by polarization in the $\pi$ channel. In contrast to the CO cases, the polarization is stronger for BaO than for MgO. 
Hence, we conclude that the change in polarization is a good measure of the adsorption strength for the AEMOs.

We find that the intramolecular CO bond length essentially remains unchanged for
upon adsorption on both MgO and BaO relative to its gas phase value of $R_{\rm CO}=1.121$~{\AA} for BEEF-vdW, while it is marginally shorter for geometries obtained with the HSE functional. This also indicates a small repulsive interaction in the $\sigma$ channel, 
which often has been observed upon CO adsorption on pure metals.~\cite{nilsson_chemical_2004}
On the other hand, the elongation of the internal NO bond is clearly visible 
for BaO: $R_{\rm NO}=1.166$~{\AA} (MgO: $R_{\rm NO}=1.146$~{\AA}) when compared to the bond of gas phase NO molecule $R_{\rm NO}=1.145$~{\AA} (BEEF-vdW functional). This observation is consistent with the picture of $\pi$-back donation to the NO molecule, which is similar to the case of NO@Ru(001) and NO@NiO(100).~\cite{staufer_interpretation_1999,rohrbach_molecular_2005}

\begin{figure*}
\includegraphics[width=\columnwidth]{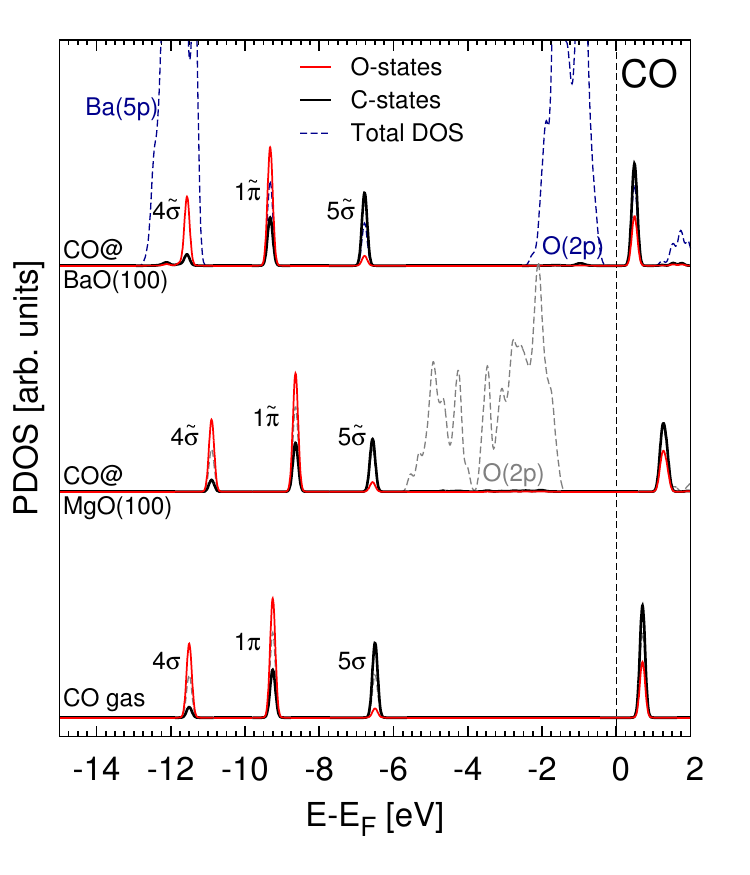}
\includegraphics[width=\columnwidth]{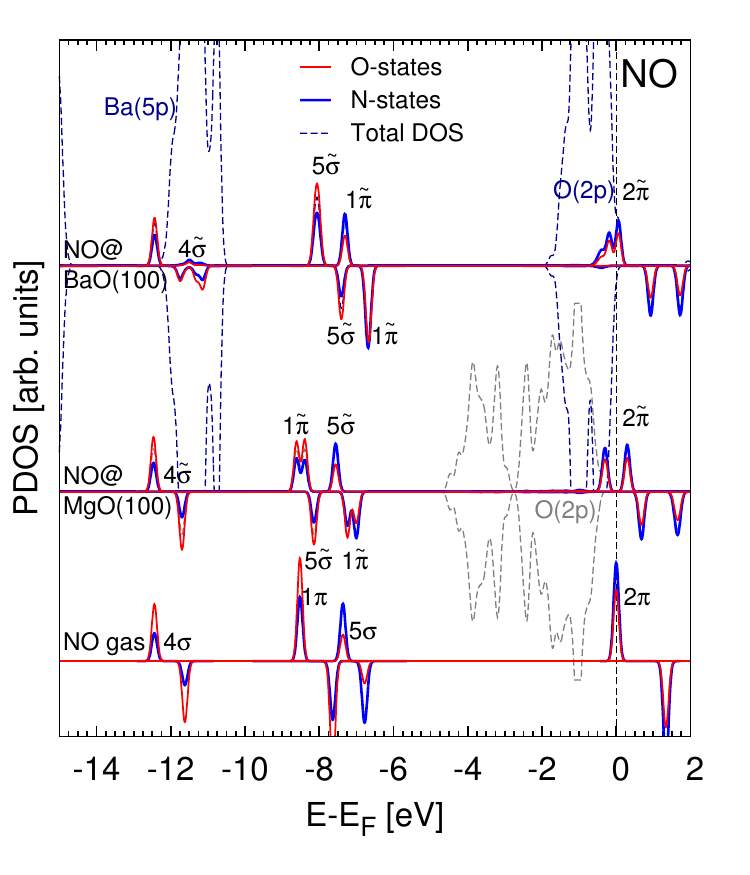}
\caption{Projected density of states of CO (left panel) and NO (right panel) adsorbed on the MgO(100) and BaO(100) surfaces compared 
to their gas-phase spectra as calculated within the BEEF-vdW functional. All spectra are aligned relative to the Fermi level $E_F$, 
except for the gas-phase spectrum of CO, which was aligned to match the position of the 5$\sigma$ peak to an experimental spectrum (adapted from Fig. 43 of Ref.~[\onlinecite{nilsson_chemical_2004}]).
For NO, the spin-minority channel is indicated as negative values. 
The atomic projections to $\sigma$ and $\pi$ contributions of molecules have been enhanced by a factor of 2 for CO and 4 for NO for clarity. 
}\label{fig:COpdos}
\end{figure*}

The projected density of states (PDOS) for CO and NO on the (100) surface of MgO and BaO are shown in Fig.~\ref{fig:COpdos}. 
For the adsorbed CO, the $4\tilde{\sigma}$, $5\tilde{\sigma}$, and $1\tilde{\pi}$ adsorbate states 
are very similar to the ones of the gas-phase CO molecule. Only a small hybridization with the states of the surface Ba($5p$) atoms is observed.
The change in intensity of the carbon $5\tilde{\sigma}$ state for CO@MgO(100) is also detected, although its value is likely underestimated 
due to the small radii employed in the atomic orbital projectors. Overall, the shifts in the position of the peaks are larger for CO@MgO(100) than for CO@BaO(100),
in agreement with the larger polarization found in the charge density differences (see Fig.~\ref{fig:COddplot}).

For adsorbed NO, the $4\tilde{\sigma}$, $5\tilde{\sigma}$, and $1\tilde{\pi}$ adsorbate states are significantly more hybridized than 
for adsorbed CO. We find that the larger tilting angle ($47\degree$) of the NO towards the MgO(100) surface leads to a lift of the degeneracy between the
$\pi_x$ and $\pi_y$ states. A small hybridization of the $2\tilde{\pi}$ state with the surfaces and a reordering of the $1\tilde{\pi}$ state relative to the $5\tilde{\sigma}$ state in the minority spin channel is also observed. For NO@BaO(100), the significantly smaller tilting angle ($17\degree$) of the NO molecule is not sufficient to lift the $\pi_{x,y}$ degeneracy. However, the $4\tilde{\sigma}$ and $2\tilde{\pi}$ states are significantly more hybridized with the states of the Ba($5p$) and O($2p$) surface atoms. 
The relative ordering of the $1\tilde{\pi}$ to $5\tilde{\sigma}$ is reversed within both spin channels. 
Again, we find overall larger hybridization and shifts in the states for NO@BaO(100) than for NO@MgO(100), which is consistent with the picture of a larger polarization observed in the density difference (see Fig.~\ref{fig:COddplot}). In other words, the differences in adsorption of CO and NO on the MgO(100) and BaO(100) surfaces can be rationalized from the electronic structure.

\subsection{Adsorption of Atomic Oxygen and Oxygen Vacancy Formation}

\begin{figure}
\includegraphics[width=\columnwidth]{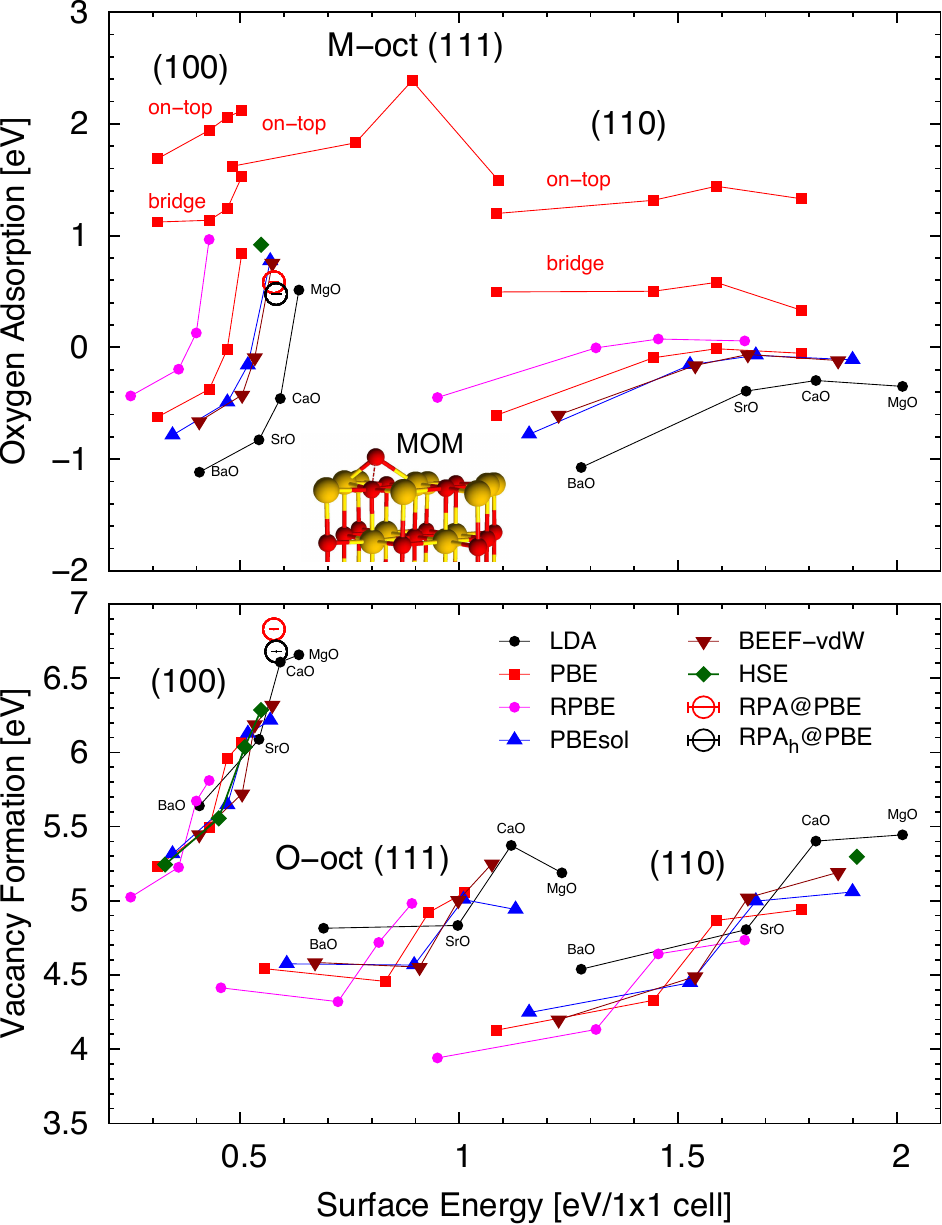}
\caption{Upper panel: Adsorption energy of atomic oxygen $E_{\rm ads.}^{\rm O}$, for adsorbate coverage $\Theta=0.25$,
on the surfaces of the AEMO series shown as function of surface energy. 
As in Fig.~\ref{fig:COads}, the reported results are for six DFT functionals and two RPA methods.
For on-top and bridge sites, only the PBE results are shown. The inset depicts the geometry of the MOM site.  
Lower panel: Same as above but for oxygen vacancy formation energy $E_{\rm f}^{\rm O}$.
}\label{fig:Oads_vs_surf}
\end{figure}

Another very important property of an oxide surface is its reactivity towards oxygen.
We have investigated oxygen adsorption energy dependence on different surface sites of the AEMO surfaces as well as the 
formation of oxygen vacancies. For the (100) and (110) surfaces, in addition to direct on-top adsorption, 
we have identified two more stable configurations: the metal-metal bridge sites as well as in 
metal-oxygen-metal (MOM) sites, in which an O$_2$-like structure is formed (shown as inset of Fig.~\ref{fig:Oads_vs_surf}). 
The energetic map of adsorption for studied surfaces of the AEMO series is shown in the upper panel of Fig.~\ref{fig:Oads_vs_surf}. 
In the lower panel of Fig.~\ref{fig:Oads_vs_surf}, we report 
the calculated oxygen vacancy formation energies, more specifically the formation energies for the neutral surface F centers.  
It is customary to define the oxygen adsorption energy $E_{\rm ads.}^{\rm O}=E_{\rm \{surf.+O ads.\}}-E_{\rm surf.}-\mu_{\rm O}$ 
and oxygen vacancy formation energy $E_{\rm f}^{\rm O}=E_{\rm surf}-E_{\rm \{surf-O\}}-\mu_{\rm O}$
relative to gas phase O$_2$ as $\mu_{\rm O}=E_{\rm DFT}({\rm O}_2)/2$.
Note, that a straightforward conversion to atomic oxygen as reference can be obtained by subtracting 
half of its experimental atomization energy (5.21~eV). An alternative definition is possible by using water and hydrogen as reference~\cite{ebensperger_first-principles_2011} as follows $\mu_{\rm O}=E_{\rm DFT}({\rm H}_2{\rm O}) - E_{\rm DFT}({\rm H}_2) +{\rm 2.506\,eV}$, which  leads to only a small constant shift ($\sim$ 0.1 eV) in observed values.  
As in the CO and NO calculations, we employed identical 4 layer slab models and $2\!\times\!2$ simulation cells. For the (111) surfaces, the oxygen adsorption energy is calculated at the on-top metal site for the M-oct termination, while the oxygen vacancy formation is calculated by oxygen removal from the O-oct terminated surface. 

We find that the adsorption at the on-top metal site on any surface is very weak, 
with adsorbed oxygen maintaining its triplet spin state ( see Fig.~\ref{fig:Oads_vs_surf}).
A stronger adsorption is found for the oxygen on the bridge sites of the (100) and (110) surfaces by about 0.6~eV and for these systems some residual spin polarization remains on the adsorbed oxygen. 
The most stable adsorption site on the (100) and (110) surfaces is the MOM site, 
which is about 1~eV more stable than the on-top site. Adsorbed oxygen at the MOM site forms a stretched O$_2$-like structure, 
which is further stabilized by the surrounding metal atoms (shown as inset of Fig.~\ref{fig:Oads_vs_surf}). 
In this case, the magnetic moment of the oxygen is fully quenched by the surface. 
The calculated oxygen vacancy formation energies are shown in the lower panel of Fig.~\ref{fig:Oads_vs_surf}. 
For all the studied oxide surfaces, we find that the final electronic configuration of the oxygen vacancy is the same as for the singlet-type neutral F-center. 

The geometric structure of the local adsorption environment plays a crucial role in the interaction between oxygen and surface.
The two surfaces with higher surface energies have also a higher number of uncoordinated bonds when compared to the (100) surface and the oxygen generally adsorbs stronger on them, as we discussed above. 
We find a similar inverse effect for the oxygen vacancy formation energy, that is the formation energy decreases for less stable surfaces 
relative to the (100) surface. However, its clear that the surface energy is only a weak indicator of the oxygen reactivity
in these systems.        

The electronic structure differences between the oxides of the AEMO series are also clearly visible. 
For a given adsorption site, the adsorption is strengthened along the AEMO series, 
similar to what we found for NO adsorption on the (100) surface. For the MOM site, this gain is as large as 1.5~eV, \textit{i.e.}, the adsorption energy difference between O@MgO(100) and O@BaO(100). 
A comparison between the density of states of O adsorbed on MgO(100) and BaO(100) (not shown) indicates a much 
larger hybridization between O$(2p)$ and BaO states.
For vacancy formation, we observe destabilization of the vacancy as the co-valency of the oxide increases. 
This is reasonable, since the extra electron pair in the vacancy gives rise to strong Madelung stabilization,~\cite{ferrari_electronic_1995} which is largest for MgO and decreases along the series.~\cite{pacchioni_measures_1993}

The adsorption calculations of O on the MOM sites for the different DFT functionals reveal the familiar ordering and energy range and bounds as observed for the NO and CO adsorption, that is, 
RPBE yielding a too weak and LDA a too strong binding. The variation in the adsorption energies is as large as 0.8~eV. Within these bounds, $E_{\rm ads.}^{\rm O}$ is ordered as $E_{\rm HSE}<E_{\rm PBE}<E_{\rm PBEsol} \approx E_{\rm BEEF-vdW}$.
The $E_{\rm f}^{\rm O}$ has a similar ordering as $E_{\rm ads.}^{\rm O}$. While we were unable to find any experimental or calculated oxygen adsorption energies, our vacancy formation energy of 6.28~eV for MgO(100) within HSE functional agrees very well with the recently published HSE value of 6.34~eV.~\cite{richter_concentration_2013}
Interestingly, the RPA results predict too strong oxygen binding oxygen adsorption and too large vacancy formation energies, which are very close to LDA results. This finding is independent of oxygen reference or hardness of oxygen pseudopotential indicating the inability of RPA to capture missing correlation under change of electron pairs 
(from triplet to singlet).~\cite{eshuis_electron_2012,ren_random-phase_2012}    

\begin{figure}
\includegraphics[width=\columnwidth]{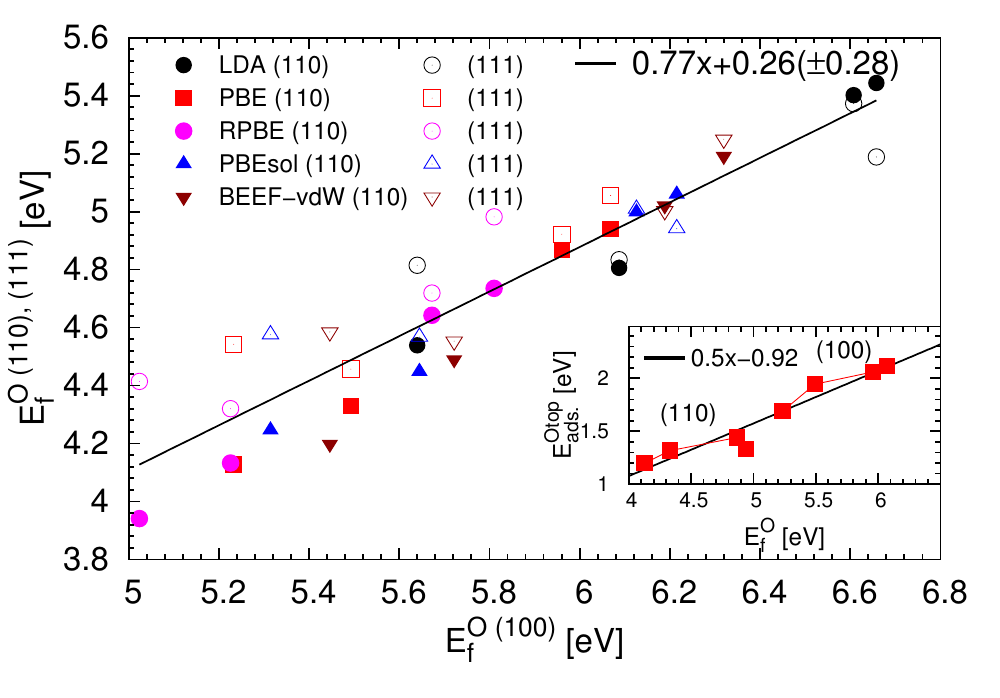}
\caption{Linear dependence of $E_{\rm f}^{\rm O}$ in (110) and O-oct (111) surfaces relative to $E_{\rm f}^{\rm O}$ of (100) surfaces.
Inset: Similar linear dependence between on-top $E_{\rm ads.}^{\rm O}$ and $E_{\rm f}^{\rm O}$ using only the 
PBE data of Fig.~\ref{fig:Oads_vs_surf}.}
\label{fig:Oads_vs_Ovac}
\end{figure}
Finally, we discuss the relations and correlations of the oxygen chemistry between different AEMOs that we have identified.
In Fig.~\ref{fig:Oads_vs_Ovac} we show that an approximate linear relation exists between the vacancy formation energies of the different AEMOs surfaces  
and the vacancy formation energies of the (100) surfaces. 
This is of utter importance, since the vacancy formation energy of a less stable surface can be predicted based on the vacancy formation of the most stable surface. 
Additionally, we also find a linear relation between on-top oxygen adsorption energies and 
surface formation energies, shown as inset of Fig.~\ref{fig:Oads_vs_Ovac}. Hence, 
a good estimate of the oxygen adsorption energetics can be established 
with the knowledge of only the oxygen vacancy formation of the (100) surfaces.

\section{Conclusions}
In conclusion, we have performed a thorough computational study of the surface chemistry
of alkaline-earth transition-metal oxides (MgO, CaO, SrO, and BaO) using different DFT functionals (LDA, PBE, RPBE, PBEsol, BEEF-vdW) and RPA methods, and benchmarked our results with existing experimental values.
The most important factors determining the surface properties of AEMOs have either an electronic or structural origin. 
We find that the electronic effects are responsible for ordering of adsorption energies within the AEMO series and structural effects are 
responsible for ordering between different facets for a given AEMO.

For surface energies, the RPA methods and the PBEsol functional provide the most accurate predictions compared to experiments. 
The best performing DFT functionals for the surface adsorption energetics on MgO(100) are found to be the BEEF-vdW functional and hybrid RPA method mostly due to direct incorporation of long-range interactions. 
A complete mapping of adsorption energetics on the AEMO surfaces using the CO and NO as probing molecules reveals a 
a stronger adsorption of NO relative to CO, which is attributed to much 
larger polarization in the $\pi$ channel of the bond between the NO and the surface. 

Finally, we establish an internal hierarchy in the oxygen chemistry of different AEMO surfaces, which include the (100), the (110), and the oxygen and metal terminated (111) surfaces. The oxygen vacancy formation energetics of all surfaces are found to be linearly correlated to the energies of the most stable (100) surfaces. In addition, we find that there is a linear relation between oxygen adsorption energies and the oxygen vacancy formation energies of the (100) surfaces. This leaves us with a scheme, at least to a first order, to approximate the reactivity of different AEMO surfaces based only on the calculated oxygen vacancy formation energies of the (100) surfaces.

\section*{Acknowledgments}
The Authors would like to thank Lars G. M. Pettersson and Anders Nilsson for useful discussions and comments.
M.B. would also like to acknowledge Mat{\' u}{\v s} Dubeck{\' y} and  Philipp Plessow for their help with RPA calculations.   
We also acknowledge the support from the project titled ``Predictive Theory of Transition Metal Oxide Catalysis:
DOE Materials Genome Project (DE-AC02-76SF00515)''. This research partially employed NERSC computational resources
under DOE Contract No. DE-AC02-05CH11231.

%
\end{document}